\documentclass[11pt]{article}
\usepackage{amsmath}
\usepackage{amssymb}
\usepackage{graphicx}
 \usepackage{indentfirst}

\renewcommand{\baselinestretch}{1.2}
  \renewcommand{\arraystretch}{1.0}
\begin{document}

 \title{Remarks on the Cryptographic Primitive of\\ Attribute-based Encryption }

 \author{Zhengjun Cao$^{1}$, \qquad Lihua Liu$^{2,*}$}
  \footnotetext{ $^1$Department of Mathematics, Shanghai University, Shanghai,
  China. \quad      $^2$Department of Mathematics, Shanghai Maritime University,   Shanghai,
  China.   $^*$\,\textsf{liulh@shmtu.edu.cn}    }

\date{}
\maketitle

\begin{quotation}
 \textbf{Abstract}.   Attribute-based encryption (ABE) which allows users to
encrypt and decrypt messages based on user attributes is a type of one-to-many encryption.
Unlike the conventional one-to-one encryption which has no intention to exclude any partners of the intended receiver from obtaining the plaintext, an ABE system tries to exclude some unintended recipients from obtaining the plaintext whether they are partners of some intended recipients. We remark that this requirement for ABE is very hard to meet.  An ABE system  cannot truly exclude some unintended recipients from decryption because some users can exchange their decryption keys  in order to maximize their own interests.  The flaw discounts the importance of the cryptographic primitive.

 \end{quotation}

\section{Introduction}
The cryptographic primitive of attribute-based encryption was introduced by Sahai and Waters \cite{SW05}.
 In the scenario, a user presents an authority with a set of credentials
that prove their right to fulfill an attribute. The authority issues a certification for the user to establish that the
user fulfills the semantic of the attribute.
This process is repeated for all attributes appropriate to
each user.  As a result, a user's identity is composed
of a set, $S$, of strings which serve as descriptive attributes
of the user. Like traditional Identity-based encryption, a
party in an ABE system only needs to know the receivers' description in order to determine their
public key.
 For example, a user's identity could
consist of attributes describing their university, department,
and job function. A party  can then specify
another set of attributes $S'$ such that a receiver can only
decrypt a message if his identity $S$ has at least $k$ attributes
in common with the set $S'$, where $k$ is a parameter set by
the system.

Attribute-based encryption has attracted much attention. Lewko, Waters, Pirretti, Goyal, and Yamada, et al \cite{AHL12,GPSW06,LOS10,LWa11,LW11,LW12,PTMW06,YAH14} studied the construction of  ABE systems. Ostrovsky,  Sahai, and   Waters \cite{OSW07} investigated some non-monotonic access structures of ABE.
 Bethencourt, Sahai,  Waters, and Goyal, et al proposed some ciphertext-policy attribute-based encryption schemes \cite{BSW07,GJP08,W11}.
 Chase and Chow  \cite{CC09,Ch07} introduced the setting of multi-authority in ABE.  Hohenberger and Waters  \cite{HW14} discussed oline/offline attribute-based encryption.
Most of these constructions use bilinear groups and some linear secret-sharing schemes as building blocks.

 Unlike a conventional one-to-one encryption, an attribute-based encryption is a type of one-to-many encryption; that is, there could
be several intended recipients that are able to decrypt a same ciphertext. Since there are many intended recipients, each recipient undertakes partial obligations to
keep the privacy of the plaintext.
An intended recipient possibly forwards the plaintext to others or shares his decryption key with others.
That is to say,  the confidentiality level in an ABE system is much lower than that in a conventional one-to-one encryption.

We here  stress that the conventional one-to-one encryption  has no intention to exclude any partner of the intended recipient from decryption. But on the contrary,
an ABE scheme tries to exclude some unintended recipients from decryption whether they are partners of some intended recipients.
In this note, we remark that some users in an ABE system can exchange their decryption keys in order to maximize their own interests, which means
that an ABE system cannot truly exclude some unintended recipients from decryption. The flaw renders the cryptographic primitive unrealistic.

\section{Different confidentiality levels}

Confidentiality is a fundamental information security objective which is a  service used to keep the content of information from all but those
authorized to have it. From the sender's point of view, in a conventional one-to-one encryption the intended recipient  undertakes the full obligations to keep the privacy of the plaintext.
However, each intended recipient in a one-to-many encryption undertakes partial obligations. Based on this observation, we classify confidentiality into two kinds, \emph{strong confidentiality} and \emph{weak confidentiality}, corresponding to \emph{full obligations} and \emph{partial obligations}, separately.
This classification will be helpful to analyze the behaviors of one intended recipient and revisit the security of different encryption models.

\section{The security requirement for one-to-one encryption revisited}

It is well known that the conventional one-to-one encryption requires that the adversary without the valid decryption key cannot recover the plaintext.
Note that the adversary here is an uncharacteristic role. The requirement does not imply that some unintended recipients cannot recover or obtain the plaintext.
In real life,  some partners of the intended recipient can obtain or recover the plaintext by the following two methods.

(1) The intended recipient, Bob, forwards the plaintext to his partner, Cindy. We refer to the Graph 1 for this case.
\begin{figure}[htbp]
\begin{minipage}[t]{.8\textwidth}
\hspace*{30mm}\includegraphics[width=0.75\textwidth]{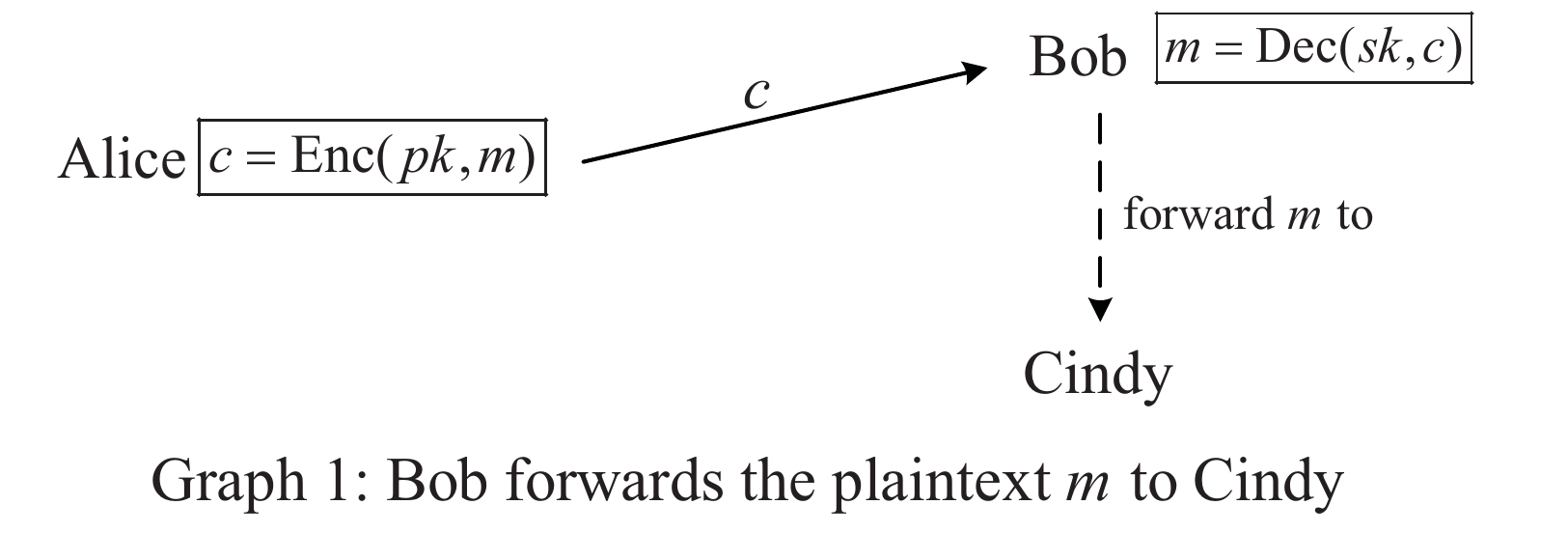}
\end{minipage}
\end{figure}

(2) The intended recipient, Bob, shares  the decryption key with his partner, Cindy. We refer to the Graph 2 for this case.
\begin{figure}[htbp]
\begin{minipage}[t]{.8\textwidth}
\hspace*{30mm} \includegraphics[width=0.75\textwidth]{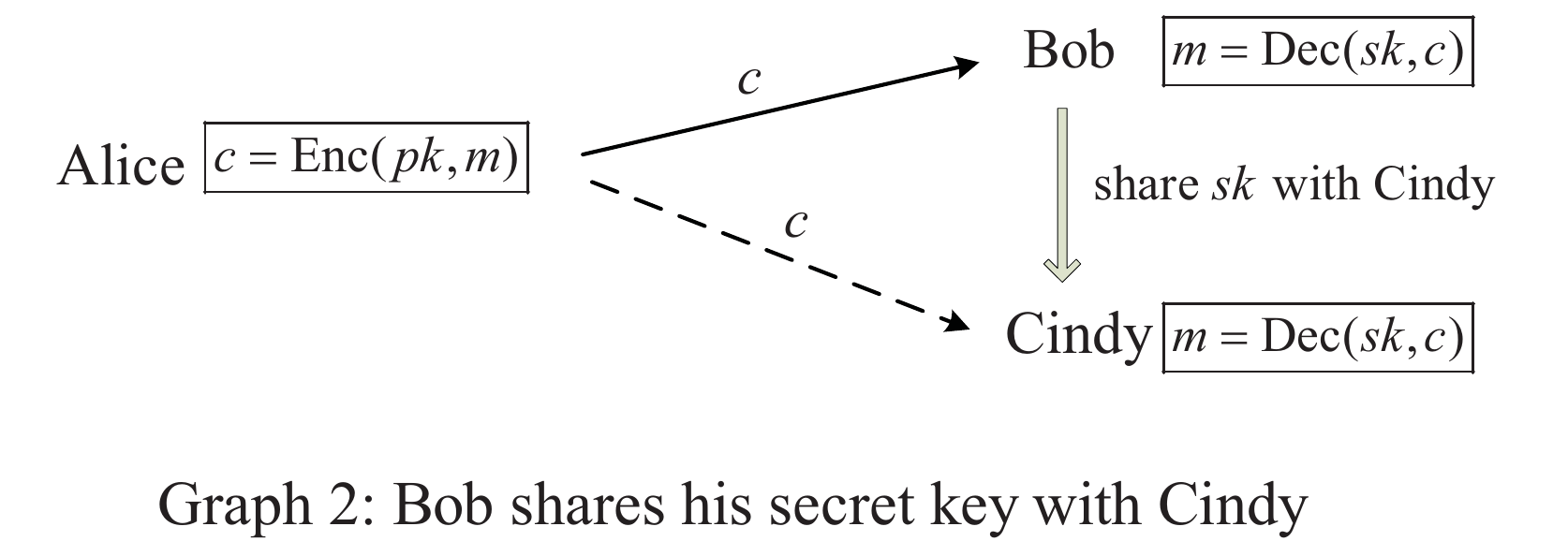}
\end{minipage}
\end{figure}

In a word,  \emph{the conventional one-to-one encryption has no intention to exclude some partners of the intended recipient from obtaining the plaintext}. This property is so obvious that it is often neglected.  However, the partnership of recipients must be taken into account when we design a one-to-many encryption system.

\section{Attribute-based encryption}

Attribute-based encryption  is claimed to be a vision of public key encryption that allows users to
encrypt and decrypt messages based on user attributes.
In the scenario, users are represented by the summation of their attributes. An encryptor will associate encrypted data with a set of attributes.
An authority will issue users different decryption  keys, where a user's decryption key is associated with an access structure over attributes and reflects
the access policy ascribed to the user. Notice that attribute-based encryption is a type of one-to-many encryption.

There are two type of attribute-based encryptions, ciphertext-policy attribute-based encryption
(CP-ABE) and key-policy attribute-based encryption (KP-ABE). In a CP-ABE system, keys are associated with sets of attributes and ciphertexts are associated with
access policies. In a KP-ABE system, the situation is reversed: keys are associated with access
policies and ciphertexts are associated with sets of attributes.
 For convenience, we only describe the definition of CP-ABE as follows.
We refer to the Graph 3 for the primitive  of ABE.

A ciphertext-policy attribute-based encryption scheme consists of the following four PPT algorithms \cite{RW12}:
\begin{itemize}
\item[] \textsf{Setup}$(1^{\lambda}) \rightarrow  (pp, msk)$: The algorithm takes the security parameter $ \lambda \in \mathbb{N} $  and outputs the public parameters $pp$ and the master secret key $msk$. Assume that the public parameters
contain a description of the attribute universe $\mathcal{U}$.

\item[]   \textsf{KeyGen}$(1^{\lambda}, pp, msk, \mathcal{S}) \rightarrow sk$: The algorithm takes  the public parameters $pp$,
the master secret key $msk$ and a set of attributes $\mathcal{S} \subseteq \mathcal{U}$.  It generates a secret key corresponding to $\mathcal{S}$.

\item[]  \textsf{Encrypt}$(1^{\lambda}, pp, m, A)\rightarrow ct$: The  algorithm takes the public parameters $pp$, a
plaintext message $m$, and an access structure $A$ on $\mathcal{U}$. It outputs the ciphertext $ct$.

\item[]
  \textsf{Decrypt}$(1^{\lambda}, pp, sk, ct) \rightarrow m$: The algorithm takes  the public parameters $pp$, a
secret key $sk$, and a ciphertext $ct$. It outputs the plaintext $m$.
\end{itemize}

A CP-ABE scheme is correct if the decryption algorithm correctly
decrypts a ciphertext of an access structure $A$ with a decryption key on $\mathcal{S}$, when $\mathcal{S}$ is an authorized set of $A$.

\begin{figure}[htbp]
\begin{minipage}[t]{.9\textwidth}
\hspace*{20mm}\includegraphics[width=0.75\textwidth]{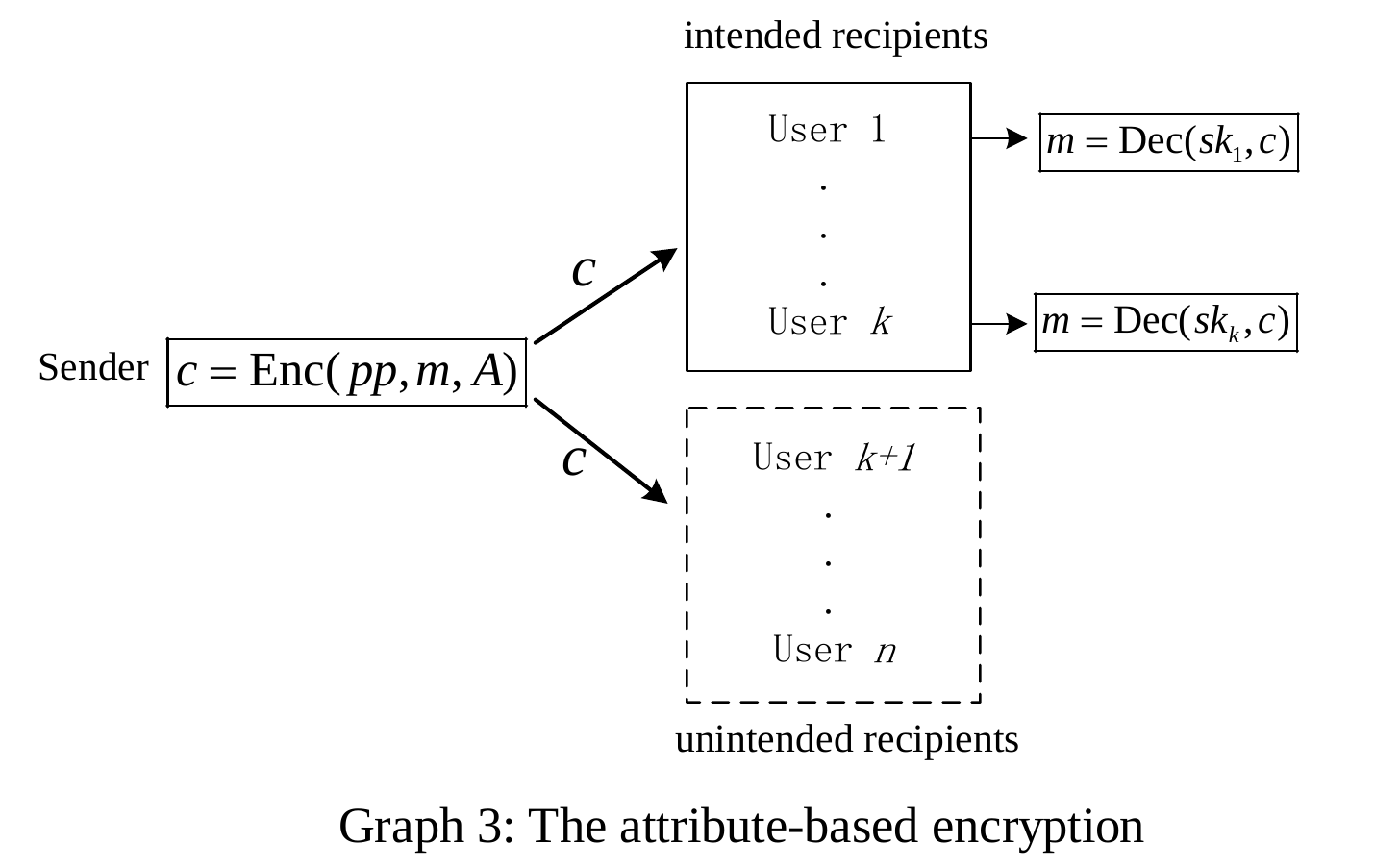}
  \end{minipage}
\end{figure}

 It is easy to find that  \emph{the attribute-based encryption tries to exclude some unintended recipients from obtaining the plaintext \underline{whether they are partners of some intended recipients}}. We shall argue that this purpose cannot be achieved.

\section{Decryption-key-sharing attack against ABE}

Most of the existing ABE schemes use bilinear groups and  some linear secret-sharing schemes as building blocks. In such an ABE system, there is an authority who is responsible for generating secret keys for all members. These secret keys are not for one-time use. They can be repeatedly invoked. Concretely, most ABE schemes have the following features:

\begin{itemize}
\item{} The secret key for each member is only used for decryption, not for signing,  because it is generated and issued by the authority. Strictly speaking,  this key has no the function of non-repudiation from a legal standpoint. Thus, it is better to call it decryption key.

\item{} One intended recipient in a communication undertakes only partial obligations to keep the privacy of the plaintext. He is more prone to reveal the plaintext to others.

\item{} Each member may become one unintended recipient in future communications. In this situation, a member is more prone to reveal his secret key to his partners if these partners are also in the same system.

\item{} In order to maximize their interests (the capability to correctly decrypt future communications),
some members can exchange their decryption keys and create alliances with as many different people in the same system as they can. For convenience, we call it decryption-key-sharing attack. We refer to the Graph 4 for this attack.
\end{itemize}

In short, \emph{the attribute-based encryption can not truly exclude some unintended recipients from decryption.}

\begin{figure}[htbp]
\begin{minipage}[t]{.9\textwidth}
\hspace*{20mm} \includegraphics[width=0.85\textwidth]{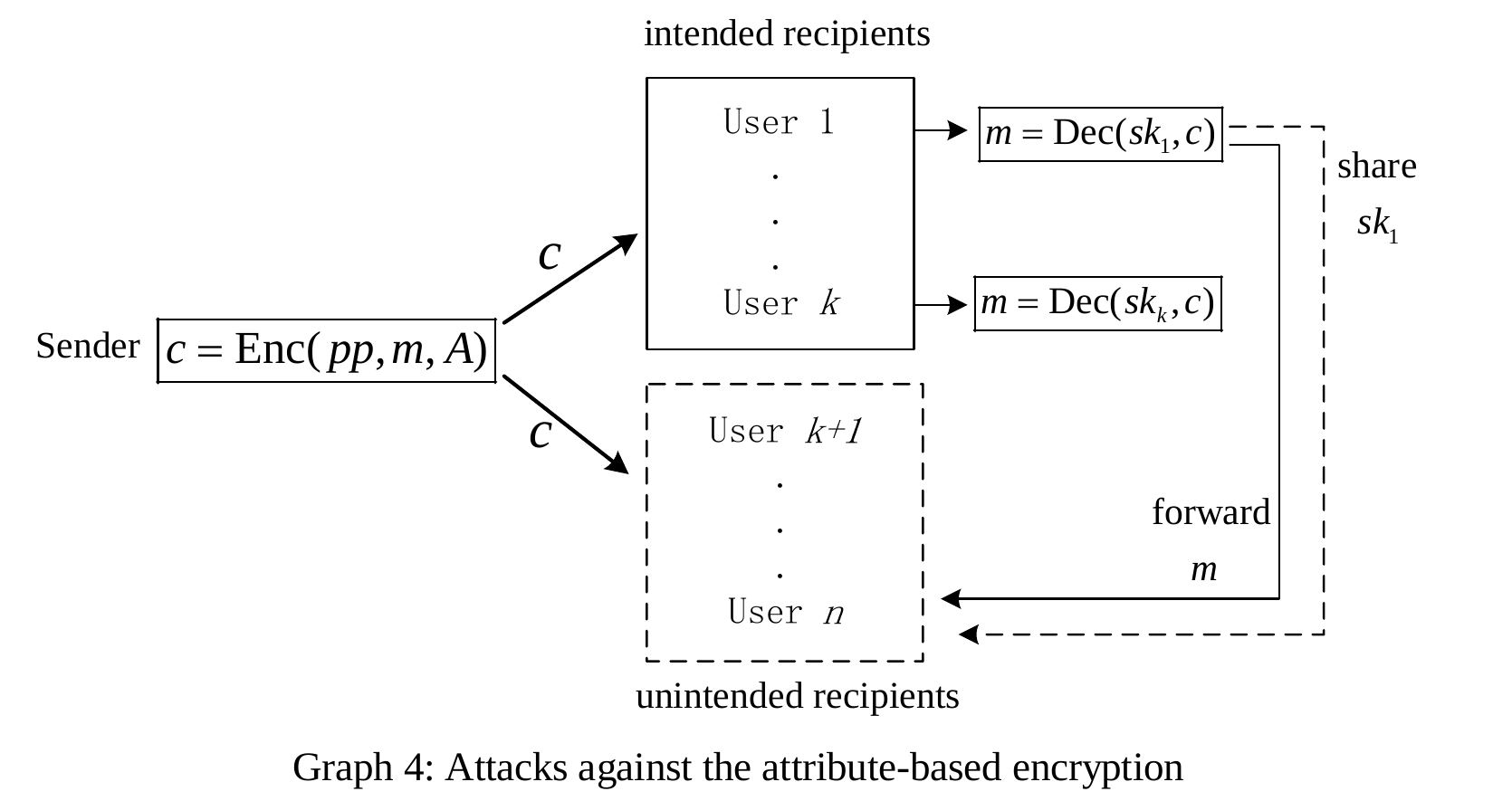}
\end{minipage}
\end{figure}

 \section{On some hypothetical applications of ABE}
It claims that
attribute-based encryption has enormous potential for providing
data security in distributed environments. We shall have a close look at the proposed examples.

\textbf{Example 1} (see \cite{PTMW06}).  A user Bob looking for
employment in the field of secure systems engineering could
place a copy of his resume in publicly accessible web space
encrypted with the attributes ``secure systems engineering"
and ``human resources manager". Only potential employers
satisfying these attributes would be able to decrypt this
information and contact Bob.

\emph{Remark 1}. We think it is better for Bob to distribute his resume through mass emails as usual.
The traditional job-hunting method could be more effective than placing the encrypted resume in publicly accessible web space.

\textbf{Example 2} (see \cite{BSW07}). Suppose that the FBI public corruption offices
in Knoxville and San Francisco are investigating
an allegation of bribery involving a San Francisco lobbyist
and a Tennessee congressman. The head FBI
agent may want to encrypt a sensitive memo so that
only personnel that have certain credentials or attributes can access it. For instance, the head agent
may specify the following access structure for accessing
this information: ((``Public Corruption Office"
AND (``Knoxville" OR ``San Francisco")) OR
(management-level $>$ 5) OR ``Name: Charlie
Eppes").  By this, the head agent could mean that the memo
should only be seen by agents who work at the public
corruption offices at Knoxville or San Francisco, FBI
officials very high up in the management chain, and a
consultant named Charlie Eppes.

\emph{Remark 2}. The bribery allegation concerning a congressman requires strong confidentiality. We do not think the primitive of ABE is appropriate to this situation because of its weak confidentiality.

\textbf{Example 3} (see \cite{SW05}).  In a computer science department, the chairperson might want to encrypt
a document to all of its systems faculty on a hiring committee. In this case it would encrypt to the
identity \{``hiring-committee",``faculty", ``systems"\}. Any user who has an identity that contains all
of these attributes could decrypt the document.

\textbf{Example 4} (see \cite{LWa11}). Suppose an administrator
needs to encrypt a junior faculty member's performance review for all senior members of the
computer science department or anyone in the dean's office. The administrator will want to
encrypt the review with the access policy (``Computer Science" AND ``Tenured") OR
``Dean's Office". In this system, only users with attributes (credentials) that match this
policy should be able to decrypt the document. The key challenge in building such systems is
to realize security against colluding users. For instance, the encrypted records should not be
accessible to a pair of unauthorized users, where one has the two credentials of ``Tenured" and ``Chemistry" and
the other one has the credential of ``Computer Science". Neither
user is actually a tenured faculty member of the Computer Science Department.

\textbf{Example 5} (see \cite{OSW07}). A university
is conducting a peer-review evaluation, where each department will be critiqued by a panel of
professors from other departments. Bob, who is a member of the panel this year from the Biology
department, will need to read (possibly sensitive) comments about other departments and assimilate
them for his written review. In an Attribute-Based Encryption system the comments will be labeled
with descriptive attributes; for example, a comment on the History department might be encrypted
with the attributes: ``History", ``year=2007", ``dept-review". In the Goyal et al scheme \cite{PTMW06},  Bob
might receive a private key for the policy ``year=2007" AND ``dept-review", which would
allow him to see all comments from this current year. However, in this setting it is important that
Bob should not be able to view comments written about his own department. Therefore, the policy
we would actually like to ascribe to Bob's key is ``year=2007" AND ``dept-review" AND
(NOT ``Biology").

\textbf{Example 6} (see \cite{HW14}). In a key-policy ABE (KP-ABE) system,
an encrypted message can be tagged with a set of attributes, such as tagging an email with the
metadata ``from: Alice", ``to: IACR board", ``subject: voting", ``date: October 1, 2012", etc. The
master authority for the system can issue private decryption keys to users including an access
policy, such as giving to Bob a decryption key that enables him to decrypt any ciphertexts that
satisfy ``to: Bob" OR (``to: IACR board" AND (January 1, 2011 $\leq$ ``date" $\leq$ December 31, 2012)).

\emph{Remark 3}. All the above four examples are contrived. They do not consider the partnership of users. For example, a user with the attributes of ``Tenured" and ``Chemistry" is very likely to be a close friend of one user with the attributes of ``Tenured" and ``Computer Science". It is a better choice for them to exchange their decryption keys in order to enhance their capabilities to decrypt future communications correctly, if they feel it is necessary. That is, the security of these examples depends on the will of users rather than some intractable assumptions.

\section{On broadcast encryption and revocation system}

Broadcast encryption is another primitive of one-to-may encryption which was formalized by Fiat and Naor \cite{FN93}.
It requires that the broadcaster encrypts a message such that a particular
set of users can decrypt the message sent over a broadcast channel.
The Fiat-Naor broadcast encryption and the works \cite{GSW00,GSY99,KRS99,S97,ST98} use a combinatorial
approach. This approach  has to right the  balance between the efficiency and the number of colluders that the system is resistant to.
Most of these schemes require that each user's decryption key is for one-time use. They have no intention to exclude some particular recipients from obtaining the plaintext.
Therefore, they are immune to decryption-key-sharing attack.

In a revocation system, a broadcaster encrypts a message such that a particular
set of revoked users cannot decrypt the message sent over a broadcast channel. In 1998, Kurosawa and Desmedt \cite{KD98} introduced a method based on polynomial interpolation for constructing revocation systems.
The subsequent revocation systems \cite{NP00,Y05} adopt this technique. In 1999, Canetti et al \cite{C99,CMN99}
developed a different method for multicast encryption. In 2001,  Naor, Naor and Lopspeich \cite{NNL01} proposed a stateless tree-based revocation scheme. Their method was subsequently improved by Halevy and Shamir \cite{HS02},  by Goodrich, Sun, and Tamassia \cite{GST04}, and by Dodis and Fazio \cite{DF02}.

At IEEE Symposium on Security and Privacy 2010, Lewko, Sahai and Waters \cite{LSW10} proposed a simple revocation  system with very small decryption keys.
  In the scheme, the authority generates all users' decryption keys which should be repeatedly used. Like most ABE schemes, the Lewko-Sahai-Waters revocation can not
   truly revoke some users because it can not resist decryption-key-sharing attack.  Note that the Goodrich-Sun-Tamassia tree-based revocation system \cite{GST04} is immune to this attack. They have stressed
that keys should be updated after each insertion or deletion (revocation) of a device. They have also specified the strategy for key update and tree rebalance.

\section{Conclusion}

  The partnership of recipients in an ABE system plays a key role in analyzing the  security of the  system which has been neglected in the past decade.
  We find an ABE system can not resist decryption-key-sharing attack.
    The flaw renders the primitive impractical.

\end{document}